\documentstyle[12pt]{article}

\title{
\begin{flushright}
{\normalsize Yaroslavl State University\\
             Preprint YARU-HE-94/09\\
             hep-ph/9411278} \\[5mm]
\end{flushright}
	Muon decays with lepton--number violation via vector leptoquark}

\author{A.A.~Gvozdev, A.V.~Kuznetsov, N.V.~Mikheev,\\
L.A.~Vassilevskaya\thanks
{E-mail addresses: phystheo@univ.uniyar.ac.ru,
phth@cnit.yaroslavl.su}\\
{\small\it Division of Theoretical Physics, Department of Physics,}\\
{\small\it Yaroslavl State University, Sovietskaya 14,}\\
{\small\it 150000 Yaroslavl, Russian Federation.}}

\date{}

\begin{document}

\maketitle

\begin{abstract}

The decays $\mu \rightarrow e \gamma$, $\mu \rightarrow e \gamma \gamma$,
and $\mu \rightarrow e e \bar e$ are analysed in the framework
of the Pati-Salam type quark-lepton symmetry
$SU(4)_V \otimes SU(2)_L \otimes G_R$ where the effects of mixing in
the quark-lepton currents are taken into account. It is shown that the
$\mu \rightarrow e \gamma \gamma$ and $\mu \rightarrow e e \bar e$ decays
via the vector leptoquark have not a GIM--type suppression, while the
$\mu \rightarrow e \gamma$ decay has. So, the specific hierarchy of the
decay probabilities could take place $\Gamma(\mu \rightarrow e e \bar e) \gg
\Gamma(\mu \rightarrow e \gamma \gamma) \gg
\Gamma(\mu \rightarrow e \gamma)$.
The existing bounds on the vector leptoquark mass
and on the mixing matrix elements, based on the data for the $\mu e$
conversion in nuclei and for the ratio of the $K_{e 2}$ and $K_{\mu 2}$
decays allow to set the upper limits on the
branching ratios at a level of $10^{-18}$
for the $\mu \rightarrow e \gamma \gamma$ decay
and at a level of $10^{-15}$ for the $\mu \rightarrow e e \bar e$ decay.

\end{abstract}

\vglue 5mm

\begin{center}
{\it Submitted to Physics Letters B}
\end{center}

\newpage

All existing experimental data in particle physics are in good agreement with
the Standard Model predictions. However, the problems exist which could not
be resolved within the SM comprehensively. A phenomenon of the fermion mixing
in the charged weak currents appears to be one of the most intriguing of them.
An effect of the mixing in the quark sector is depicted by the
Cabibbo--Kobayashi--Maskawa unitary $(3 \times 3)$ matrix $V_{ij}$.
It is measured
with rather high accuracy~\cite{PDG}, and the information on mixing parameters
is being permanently improved. It would be natural to expect of the analogous
mixing phenomenon to take place in the lepton sector, provided the neutrino
mass spectrum is not degenerated. The neutrino oscillation
experiments~\cite{Bah} are the main source of information on the lepton
mixing parameters.
The lepton--number violating decays, such as $\mu \rightarrow e \gamma$,
$\mu \rightarrow e \gamma \gamma$, $\mu \rightarrow e e \bar e$~\cite{Wal}
are also under the intensive experimental searches. Let us point out,
however, that these decay modes are strongly suppressed in the SM due to
the well--known GIM cancellation by the factor

\begin{equation}
\left (\frac{m_{\nu}}{m_W} \right )^4 \; \sim \; 10^{-39} \cdot
\left (\frac{m_{\nu}}{20 \,eV} \right )^4 ,
\label{eq:GIM}
\end{equation}

\noindent see, e.g. Refs.~\cite{Bil,VGM}.
Due to the smallness of neutrino mass, a conclusion is forced for the
processes having this width to be
unobservable in a laboratory. If an existence of the fourth generation is
assumed, the neutral lepton of it must be heavy, $m_{L^0} \, > \, m_Z/2$,
and the suppression of the type of Eq.~(\ref{eq:GIM}) disappears. It should be
noted that another kind of suppression could arise in this case by the small
mixing angles. Really, a noticeable mixing of the leptons of the 4th
generation with the 1st and the 2nd has to violate the $\mu-\beta$
universality and to cause an effect of the non-orthogonality of the
phenomenological electron and muon neutrinos~\cite{Lip}.

There exists an alternative possibility of the Standard Model extension
where the above--mentioned rare muon decays could arise without the GIM
suppression. It is the Minimal Quark--Lepton Symmetry of the Pati--Salam
type based on the gauge group $SU(4)_V \otimes SU(2)_L \otimes G_R$ with
lepton number as the fourth color~\cite{PSm}. In the recent parers~\cite{KM}
the possible low--energy manifestations of this symmetry were analysed.
There exist the most exotic objects as the fractionally--charged and colored
gauge $X$ bosons named leptoquarks which cause the interconversions of
quarks and leptons. As was shown in~\cite{KM},
a new type of mixing in the quark-lepton current interactions with
leptoquarks has to be taken into consideration.
An additional arbitrariness of the mixing parameters could allow, in principle,
to decrease noticeably the lower bound on the vector leptoquark mass
$M_X$ originated from the rare $\pi$ and $K$ decays~\cite{PDG}.
The only mixing independent bound emerging from the
cosmological limit on the $\pi^0 \rightarrow \nu \bar{\nu}$ decay
width~\cite{Lam} is $M_X > 18~TeV$.

Let us investigate the contribution of the leptoquark interactions to the
decays $\mu \rightarrow e \gamma$, $\mu \rightarrow e \gamma
\gamma$ and $\mu \rightarrow e e \bar e$. The corresponding part of the
Lagrangian of the $down$--type fermion interaction with the
leptoquark has the form

\begin{equation}
{\cal L}_X \, = \, \frac{g_S(M_X)}{\sqrt 2} \big [
{\cal D}_{\ell n}
\big ( \bar \ell \gamma_{\alpha} d^c_n \big ) X^c_{\alpha} + h.c. \big ] \,,
\label{eq:LX}
\end{equation}

\noindent where $c$ is the $SU(3)$ color index, the indices $\ell$ and $n$
correspond to the $down$--fermions, namely, charged leptons $\ell = e, \mu,
\tau$ and quarks $n = d,s,b$. The constant \, $g_S(M_X)$ \, can be expressed
in terms of the strong coupling constant \, $\alpha_S$ \, at the leptoquark
mass scale $M_X, \quad g_S^2(M_X)/4 \pi = \alpha_S(M_X)$.

If the momentum transferred is \, $q \ll M_X$, \, then the Lagrangian
{}~(\ref{eq:LX}) leads to an effective four-fermion interaction of the
quark--lepton vector currents. By using the Fiertz
transformation, lepton and quark currents of the scalar,
pseudoscalar, vector and axial-vector types may be separated in the
effective Lagrangian. Let us note that the construction of the effective
lepton-quark interaction Lagrangian requires taking account of the QCD
corrections estimated by known techniques~\cite{Vai,Vys}.
In our case the leading log approximation $ln(M_X/\mu_0) \gg 1$ with
$\mu_0 \sim 1~GeV$ to be the typical hadronic scale is quite applicable.
Then the QCD correction amounts to the appearance of the magnifying factor
$Q(\mu_0)$ at the scalar and pseudoscalar terms

\begin{equation}
Q(\mu_0) \, = \, \left ( \frac{\alpha_S(\mu_0)}
{\alpha_S(M_X)} \right )^{4/\bar b} \, .
\label{eq:Qmu}
\end{equation}

\noindent Here $\alpha_S(\mu_0)$ is the effective strong coupling constant
at the hadron mass $ \, \mu_0 \,$ scale,
$\; \bar b \, = \, 11 \, - \, \frac {2}{3} \bar n_f, \; \bar n_f $
is the averaged number of the quark
flavors at the scales $\mu_0^2 \le q^2 \le M_X^2$. If the condition
$M_X^2 \gg m_t^2$ is valid, then we have $\, \bar n_f \, \simeq \, 6$,
and $\bar b \, \simeq \, 7$.

A part of the effective Lagrangian we are interested in, providing the
lepton-number nonconserving transitions, has the form

\begin{eqnarray}
{\cal L}_{eff} \; = \; - \frac{2 \pi \alpha_S(M_X)}{M_X^2} \;
{\cal D}_{\ell n} {\cal D}^*_{\ell' n'}
\; \big \{ \frac{1}{2} \, \big ( \bar \ell \gamma_{\alpha} \gamma_5 \ell'
\big ) \big ( \bar d_{n'} \gamma_{\alpha} \gamma_5 d_n \big )
\nonumber \\
+ \; (\gamma_5 \rightarrow 1) \; + \; Q(\mu_0)
\; \big [ \big ( \bar \ell \gamma_5 \ell' \big )
\big ( \bar d_{n'} \gamma_5 d_n \big )
\; - \; (\gamma_5 \rightarrow 1)
\big ] \big \}.
\label{eq:Lef}
\end{eqnarray}

\medskip

\noindent Each of the processes under consideration is described by a number
of the one-loop Feynman diagrams with virtual $d, s, b$ quarks. The diagrams
giving the main contributions to the amplitudes are shown in Figs.1,2,3.
It is worthwhile to divide the range of integration over the virtual
momentum $k$ in the loops of Figs.1,2,3 into two parts, taking for the
dividing point some scale $\Lambda_0$ with the perturbative QCD be applicable
above it. It seems reasonable to take
$\Lambda_0 \sim (2 \div 3) \Lambda_{QCD}$ if we intend to make
the estimations in order of magnitude only. Then the decay amplitudes
could be represented in the form ${\cal M}  = \Delta {\cal M}^{LD} +
\Delta {\cal M}^{SD}$. Here $\Delta {\cal M}^{SD}$ is the short-distance
contribution corresponding to the range of a big virtual momenta
$k > \Lambda_0$ where the free-quark approximation is quite applicable. On
the other hand, in an estimation of the long-distance contribution
$\Delta {\cal M}^{LD}$ ($k < \Lambda_0$) where the perturbative QCD does
not work, we use the pole--dominance model.

As the analyses of the radiative muon decays show, the two--photon decay
dominates the one--photon decay in the model considered. Really, as the
squared
momentum transferred is $q^2 \simeq m^2_{\mu}$, the main contribution to the
$\mu \rightarrow e \gamma \gamma$ decay amplitude $\Delta {\cal M}^{LD}_{2
\gamma}$ is obviously originated from the virtual $\pi^0$ meson, see Fig.4,
being rather close to the mass--shell. It is sufficient in this case to
consider the pseudoscalar term only in the effective Lagrangian~(\ref{eq:Lef}).
In these approximations the long--distance contribution to the
$\mu \rightarrow e \gamma \gamma$ decay amplitude is

\begin{eqnarray}
\Delta {\cal M}^{LD}_{2 \gamma} \, \simeq \, - \, \frac{i \alpha \,
\alpha_S(M_X)}{2 \, M^2_X} \,
{\cal D}_{e d} {\cal D}^*_{\mu d} \;
\frac{Q(\mu_0)}{m_d(\mu_0)} \,
\frac{1}{1-q^2/m^2_{\pi}} \nonumber\\[3mm]
\times \big ( \bar e \gamma_5 \mu \big ) \; f_{1 \rho \sigma} \,
{\tilde f}_{2 \sigma \rho},
\label{eq:MLD}
\end{eqnarray}

\noindent where $\alpha$ is the fine structure constant, $f_{\rho \sigma} =
k_{\rho} \epsilon_{\sigma} - k_{\sigma} \epsilon_{\rho}$ is the Fourier
transform of the photon field tensor, ${\tilde f}_{\sigma \rho} = \frac{1}{2}
e_{\sigma \rho \alpha \beta} \, f_{\alpha \beta}$ is the dual tensor,
$q = k_1+k_2$ is the total momentum 4--vector of the photon pair,
$\, m_{d}(\mu_0) \,$ is the running mass of the $d$ quark at the $\mu_0$
scale. Let us note that the ratio $Q(\mu)/m(\mu)$
is the renormalization group invariant, since the $Q(\mu)$
function~(\ref{eq:Qmu}) determines also the law of the quark mass running.
To the $\mu_0 \simeq 1~GeV$ scale there correspond the well-known quark
current masses $m_u \simeq 4~MeV, m_d \simeq 7~MeV$ and
$m_s \simeq 150~MeV$, see e.g. Refs.~\cite{Gas,W}.

The pole approximation gives zero result for the analogous contribution
$\Delta {\cal M}^{LD}_{1 \gamma}$ to the $\mu \rightarrow e \gamma $ decay
amplitude, because no intermediate meson state exists to pass into a real
photon.

It is interesting to note that the short--distance contribution
$\Delta {\cal M}^{SD}_{1 \gamma}$ to the $\mu \rightarrow e \gamma$ decay
is also strongly suppressed as compared with the corresponding contribution
to the $\mu \rightarrow e \gamma \gamma$ decay. It can be easily understood
from the following qualitative treatment. If the $\mu \rightarrow e \gamma$
process was considered in the local limit of the quark--lepton interaction,
see Fig.1, then the quark loop would have only one external momentum, namely,
the photon momentum $q$, and the gauge invariant amplitude with the real
photon could not be constructed. Additional momenta of the external particles
could appear if the non--local effects in the quark--lepton interaction were
only considered. However, an extra factor of suppression
$\sim \left( m_b / M_X \right)^{2}$ inevitably arises in the amplitude in
this case analogously to the well--known GIM suppression, and in order of
magnitude we have

\begin{equation}
\frac{\Gamma(\mu\to e\gamma\gamma)}{\Gamma(\mu\to e\gamma)} \; \sim \;
\frac{\alpha}{\pi} \, \left( \frac{M_X}{m_b} \right) ^{4} \; \gg \; 1.
\label{eq:GIM1}
\end{equation}

\noindent The exact calculation does confirm this qualitative analysis.

When the short--distance contribution to the $\mu \rightarrow e \gamma \gamma$
decay amplitude is calculated, the vector part of the effective Lagrangian
{}~(\ref{eq:Lef}) does not work due to the Furry theorem. As the analysis
shows,
the scalar and pseudoscalar parts dominate and give the equal contributions.
We obtain

\begin{eqnarray}
\Delta {\cal M}^{SD}_{2 \gamma} \, \simeq \, \frac{\alpha \, \alpha_S(M_X)}
{3 \, M^2_X} \,
{\cal D}_{e b} {\cal D}^*_{\mu b} \;
\frac{Q(\mu_0)}{\Lambda_0} \, \big [
\big ( \bar e \mu \big ) \; f_{1 \rho \sigma} \,
f_{2 \sigma \rho} \;  \nonumber\\[3mm]
- \; i \big ( \bar e \gamma_5 \mu \big ) \; f_{1 \rho \sigma} \,
{\tilde f}_{2 \sigma \rho} \big ].
\label{eq:MSD}
\end{eqnarray}

\noindent In general, a relative sign of the amplitudes~(\ref{eq:MLD})
and~(\ref{eq:MSD}) couldn't be established in the approach used. When the
estimation in order of magnitude is performed, the interference term of
Eq.~(\ref{eq:MLD}) and of the pseudoscalar part of Eq.~(\ref{eq:MSD}) can be
omitted for simplicity and the separate contributions of the long and short
distances to the $\mu \rightarrow e \gamma \gamma$ decay branch can be found

\begin{eqnarray}
Br(\mu \rightarrow e \gamma \gamma)^{LD} \simeq
\frac{3 \, \alpha^2 \, \alpha_S^2(M_X)}{16} \;
\left ( \frac{m_{\mu} \, Q(\mu_0)}{m_d(\mu_0)} \right )^2
\nonumber\\[3mm]
\times F \left (\frac{m_{\mu}^2}{m_{\pi}^2} \right )
\left (\frac{|{\cal D}_{e d} {\cal D}^*_{\mu d}|}
{G_F \, M^2_X} \right )^2,
\label{eq:BLD}
\end{eqnarray}

\begin{equation}
F(a) =
\frac {1}{a^{2}} \, \left( \frac{1}{3} - \frac{4}{a} + \frac{4}{a^{2}} \right)
+ \frac{2}{a^{3}} \, \left( 1- \frac{3}{a} + \frac{2}{a^{2}} \right) \,
\ln{(1-a)},
\nonumber
\end{equation}

\begin{equation}
Br(\mu \rightarrow e \gamma \gamma)^{SD} \simeq
\frac{\alpha^2 \, \alpha_S^2(M_X)}{180} \;
\left ( \frac{m_{\mu} \, Q(\mu_0)}{\Lambda_0} \right )^2 \;
\left (\frac{|{\cal D}_{e b} {\cal D}^*_{\mu b}|}
{G_F \, M^2_X} \right )^2,
\label{eq:BSD}
\end{equation}

\noindent where $G_F$ is the Fermi coupling constant. The magnitude
{}~(\ref{eq:BLD}) could be estimated from the experimental data on the $\mu -
e$
conversion in nuclei~\cite{Ah} being also possible due to the leptoquark
exchange. The bound obtained in Ref.~\cite{KM} is

\begin{equation}
\frac{M_X}{|{\cal D}_{e d} {\cal D}^*_{\mu d}|^{1/2}} \, > \,
670~TeV.
\label{eq:670}
\end{equation}

Considering a rather slow increase of the running coupling constant
$\alpha_S$ with energy and assuming $\alpha_S(M_X) \sim \alpha_S(100~TeV)
= 0.063$ we obtain the following numerical estimation for the $Br^{LD}$

\begin{equation}
Br(\mu \rightarrow e \gamma \gamma)^{LD} \; < \; 1.4 \cdot 10^{-19}.
\label{eq:BrLD}
\end{equation}

Given the bound~(\ref{eq:670}) an upper limit on the combination of the
model parameters  $|{\cal D}_{e b} {\cal D}^*_{\mu b}|/M^2_X$
involved into~(\ref{eq:BSD}) could be obtained assuming that the $\tau$
lepton is associated mainly with the $d$ quark. As was shown in Ref.~\cite{VW},
the experimental data on the ratio $R_{e/\mu} = \Gamma(K^+ \to e^+ \nu)/
\Gamma(K^+ \to \mu^+ \nu)$~\cite{PDG} gives the most stringent constraint
on the model parameters in this case. The calculation of the leptoquark
contribution to it is such as for the ratio of the $\pi_{\ell 2}$
decay probabilities~\cite{KM}. As a result the bound is
$M_X/|{\cal D}_{e s}| > 55 \, TeV$. Taking into account the unitarity of the
$\cal D$ matrix and the other limits on its elements, see Ref.~\cite{KM},
one obtains

\begin{equation}
\frac{M_X}{|{\cal D}_{e b} {\cal D}^*_{\mu b}|^{1/2}} \, > \,
55~TeV.
\label{eq:55}
\end{equation}

\noindent Finally we get

\begin{equation}
Br(\mu \rightarrow e \gamma \gamma)^{SD} \; < \; 1.0 \cdot 10^{-18}.
\label{eq:BrSD}
\end{equation}

A similar analysis of the $\mu \to e e \bar e$ decay shows that the
short--distance contribution also dominates there as in the
$\mu \rightarrow e \gamma \gamma$ decay. An amplitude of the process,
see Fig.3, could be represented in the form

\begin{eqnarray}
{\cal M}_{3e} \, \simeq \,
\Delta {\cal M}^{SD}_{3e} \, \simeq \, - \, \frac{2 \alpha \, \alpha_S(M_X)}
{3 \, M^2_X} \,
{\cal D}_{e b} {\cal D}^*_{\mu b} \;
\ln{\frac{m_b}{\Lambda_0}} \nonumber\\[3mm]
\times \big [ \big ( \bar e_1 \gamma_{\alpha } \mu \big ) \,
\big ( \bar e_2 \gamma_{\alpha } e_3 \big ) \, - \,
(1 \leftrightarrow 2) \big ].
\label{eq:M3e}
\end{eqnarray}

\noindent The $\mu \to e e \bar e$ decay branch is

\begin{equation}
Br(\mu \rightarrow e e \bar e) \simeq
\frac{2 \alpha^2 \, \alpha_S^2(M_X)}{3} \;
\left ( \ln {\frac{m_b}{\Lambda_0}} \right )^2 \;
\left (\frac{|{\cal D}_{e b} {\cal D}^*_{\mu b}|}
{G_F \, M^2_X} \right )^2.
\label{eq:B3e}
\end{equation}

\noindent Within the above restrictions on the model parameters we obtain

\begin{equation}
Br(\mu \rightarrow e e \bar e)  \; < \; 1.0 \cdot 10^{-15}.
\label{eq:Br3e}
\end{equation}

The $\mu \to e e \bar e$ decay via the vector leptoquark was also considered
in Ref.~\cite{Dav} within the model independent approach, but the diagram
giving the main contribution to the amplitude, see Fig.3, was not analysed
there.

In summary, the minimal quark--lepton symmetry
$SU_V (4) \otimes SU_L (2) \otimes G_R$ with taking account of the mixing
in the quark--lepton currents leads to some interesting predictions about
the rare muon decays with a lepton number nonconservation:

{\it i)} a peculiar hierarchy of the decay probabilities could take place

\begin{equation}
\Gamma(\mu \rightarrow e e \bar e) \gg
\Gamma(\mu \rightarrow e \gamma \gamma) \gg
\Gamma(\mu \rightarrow e \gamma),
\end{equation}

\noindent see the estimations~(\ref{eq:GIM1}),
(\ref{eq:BrLD}), (\ref{eq:BrSD}), (\ref{eq:Br3e});

{\it ii)} the branches of the considered decays do not depend on the
neutrino masses. That is to say that these decays are possible even though
the neutrino mass spectrum is degenerated, e.g. all the neutrinos are
massless.

Although the predicted values of the branches of the $\mu \to e \gamma
\gamma$ and $\mu \rightarrow e e \bar e$ decays, see Eqs.~(\ref{eq:BrLD}),
{}~(\ref{eq:BrSD}), ~(\ref{eq:Br3e}), are essentially less then the existing
experimental limits

$Br(\mu \rightarrow e \gamma \gamma)_{exp} < 7.2 \cdot 10^{-11}$~\cite{Bol},

$Br(\mu \rightarrow e e \bar e)_{exp} < 1.0 \cdot 10^{-12}$~\cite{3e},

\noindent they are not as small as the Standard Model predictions, and a hope
for their observation in the future still remains.

In our opinion, the results obtained could be of interest for the discussions
of the prospects of further searches for the $\mu \rightarrow e \gamma$,
$\mu \rightarrow e \gamma \gamma$, and $\mu \rightarrow e e \bar e$ decays.

\medskip

The authors are grateful to
L.B.~Okun, V.A.~Rubakov, K.A.~Ter-Mar\-ti\-ro\-sian, and A.D.~Smirnov
for interesting discussions.

The research described in this publication was made possible in part by
Grant N RO3000 from the International Science Foundation.

\newpage

\begin{minipage}[t]{200mm}

\unitlength=0.75mm
\special{em:linewidth 0.4pt}
\linethickness{0.4pt}

%%%%%%%%%%%% Fig.1 %%%%%%%%%%%%%%%%%%%%%

\begin{picture}(50.00,57.00)(0,0)
\put(18.00,30.00){\vector(1,0){1.00}}
\put(42.00,30.00){\vector(1,0){1.00}}
\put(21.50,39.00){\vector(0,1){1.00}}
\put(38.50,39.00){\vector(0,-1){1.00}}
\put(10.00,30.00){\line(1,0){40.00}}
\put(30.00,39.00){\circle{16.00}}
\put(30.00,30.00){\circle*{1.70}}
\put(35.00,55.00){\makebox(0,0)[cc]{$\gamma$}}
\put(47.00,40.00){\makebox(0,0)[cc]{$d,s,b$}}
\put(18.00,25.00){\makebox(0,0)[cc]{$\mu$}}
\put(42.00,25.00){\makebox(0,0)[cc]{$e$}}
\put(30.00,10.00){\makebox(0,0)[cc]{\large Fig.~1.}}
\put(30.00,49.00){\oval(3.00,3.00)[r]}
\put(30.00,52.00){\oval(3.00,3.00)[l]}
\put(30.00,55.00){\oval(3.00,3.00)[r]}
\put(30.00,58.00){\oval(3.00,3.00)[l]}
\put(30.00,61.00){\oval(3.00,3.00)[r]}
\put(30.00,64.00){\oval(3.00,3.00)[l]}
\put(30.00,47.50){\circle*{1.20}}
\end{picture}

%%%%%%%%%%%% end Fig.1 %%%%%%%%%%%%%%%%%%%%%

%%%%%%%%%%%% Fig.2 %%%%%%%%%%%%%%%%%%%%%

\begin{picture}(100.00,57.00)(-90,-57)
\put(18.00,30.00){\vector(1,0){1.00}}
\put(42.00,30.00){\vector(1,0){1.00}}
\put(21.50,39.00){\vector(0,1){1.00}}
\put(38.50,39.00){\vector(0,-1){1.00}}
\put(10.00,30.00){\line(1,0){40.00}}
\put(30.00,39.00){\circle{16.00}}
\put(30.00,30.00){\circle*{1.70}}
\put(40.00,58.00){\makebox(0,0)[cc]{$\gamma_2$}}
\put(21.00,58.00){\makebox(0,0)[cc]{$\gamma_1$}}
\put(47.00,40.00){\makebox(0,0)[cc]{$d,s,b$}}
\put(18.00,25.00){\makebox(0,0)[cc]{$\mu$}}
\put(42.00,25.00){\makebox(0,0)[cc]{$e$}}
\put(50.00,10.00){\makebox(0,0)[cc]{\large Fig.~2.}}
\put(34.00,48.00){\oval(3.00,3.00)[r]}
\put(34.00,51.00){\oval(3.00,3.00)[l]}
\put(34.00,54.00){\oval(3.00,3.00)[r]}
\put(34.00,57.00){\oval(3.00,3.00)[l]}
\put(34.00,60.00){\oval(3.00,3.00)[r]}
\put(34.00,63.00){\oval(3.00,3.00)[l]}
\put(34.00,46.50){\circle*{1.20}}
\put(26.00,48.00){\oval(3.00,3.00)[l]}
\put(26.00,51.00){\oval(3.00,3.00)[r]}
\put(26.00,54.00){\oval(3.00,3.00)[l]}
\put(26.00,57.00){\oval(3.00,3.00)[r]}
\put(26.00,60.00){\oval(3.00,3.00)[l]}
\put(26.00,63.00){\oval(3.00,3.00)[r]}
\put(26.00,46.50){\circle*{1.20}}
\put(78.00,45.00){\makebox(0,0)[cc]{$\;+\quad(\gamma_1\leftrightarrow
\gamma_2)$}}
\end{picture}

%%%%%%%%%%%% end Fig.2 %%%%%%%%%%%%%%%%%%%%%

%%%%%%%%%%%% Fig.3 %%%%%%%%%%%%%%%%%%%%%

\begin{picture}(100.00,80.00)(0,-30)
\put(18.00,30.00){\vector(1,0){1.00}}
\put(42.00,30.00){\vector(1,0){1.00}}
\put(21.50,39.00){\vector(0,1){1.00}}
\put(38.50,39.00){\vector(0,-1){1.00}}
\put(10.00,30.00){\line(1,0){40.00}}
\put(30.00,39.00){\circle{16.00}}
\put(30.00,30.00){\circle*{1.70}}
\put(25.00,55.00){\makebox(0,0)[cc]{$\gamma^*$}}
\put(47.00,40.00){\makebox(0,0)[cc]{$d,s,b$}}
\put(18.00,25.00){\makebox(0,0)[cc]{$\mu$}}
\put(42.00,25.00){\makebox(0,0)[cc]{$e_1$}}
\put(43.00,63.00){\makebox(0,0)[cc]{$e_2$}}
\put(38.00,74.00){\makebox(0,0)[cc]{$e_3$}}
\put(50.00,10.00){\makebox(0,0)[cc]{\large Fig.~3.}}
\put(77.00,45.00){\makebox(0,0)[cc]{$\;-\quad(e_1\leftrightarrow e_2)$}}
\put(30.00,49.00){\oval(3.00,3.00)[r]}
\put(30.00,52.00){\oval(3.00,3.00)[l]}
\put(30.00,55.00){\oval(3.00,3.00)[r]}
\put(30.00,58.00){\oval(3.00,3.00)[l]}
\put(30.00,47.50){\circle*{1.20}}
\put(30.00,59.50){\circle*{1.20}}
\put(30.00,59.50){\line(1,0){20.00}}
\put(30.00,59.50){\line(1,1){15.00}}
\put(42.00,59.50){\vector(1,0){1.00}}
\put(39.00,68.50){\vector(-1,-1){1.00}}
\end{picture}

%%%%%%%%%%%% end Fig.3 %%%%%%%%%%%%%%%%%%%%%

%%%%%%%%%%%% Fig.4 %%%%%%%%%%%%%%%%%%%%%

\begin{picture}(70.00,80.00)(-110,-110)
\put(18.00,30.00){\vector(1,0){1.00}}
\put(42.00,30.00){\vector(1,0){1.00}}
\put(10.00,30.00){\line(1,0){40.00}}
\put(30.00,30.00){\circle*{1.70}}
\multiput(29.30,30.00)(0.00,5.00){4}{\line(0,1){3.00}}
\multiput(30.70,30.00)(0.00,5.00){4}{\line(0,1){3.00}}
\put(18.00,25.00){\makebox(0,0)[cc]{$\mu$}}
\put(42.00,25.00){\makebox(0,0)[cc]{$e$}}
\put(35.00,40.00){\makebox(0,0)[cc]{$\pi^0$}}
\put(13.00,55.00){\makebox(0,0)[cc]{$\gamma_1$}}
\put(47.00,55.00){\makebox(0,0)[cc]{$\gamma_2$}}
\put(30.00,10.00){\makebox(0,0)[cc]{\large Fig.~4.}}
\put(30.00,48.00){\circle*{1.70}}
\put(30.00,49.50){\oval(3.00,3.00)[b]}
\put(33.00,49.50){\oval(3.00,3.00)[lt]}
\put(33.00,52.50){\oval(3.00,3.00)[rb]}
\put(36.00,52.50){\oval(3.00,3.00)[lt]}
\put(36.00,55.50){\oval(3.00,3.00)[rb]}
\put(39.00,55.50){\oval(3.00,3.00)[lt]}
\put(39.00,58.50){\oval(3.00,3.00)[rb]}
\put(42.00,58.50){\oval(3.00,3.00)[lt]}
\put(42.00,61.50){\oval(3.00,3.00)[rb]}
\put(45.00,61.50){\oval(3.00,3.00)[lt]}
\put(27.00,49.50){\oval(3.00,3.00)[rt]}
\put(27.00,52.50){\oval(3.00,3.00)[lb]}
\put(24.00,52.50){\oval(3.00,3.00)[rt]}
\put(24.00,55.50){\oval(3.00,3.00)[lb]}
\put(21.00,55.50){\oval(3.00,3.00)[rt]}
\put(21.00,58.50){\oval(3.00,3.00)[lb]}
\put(18.00,58.50){\oval(3.00,3.00)[rt]}
\put(18.00,61.50){\oval(3.00,3.00)[lb]}
\put(15.00,61.50){\oval(3.00,3.00)[rt]}
\end{picture}

%%%%%%%%%%%% end Fig.4 %%%%%%%%%%%%%%%%%%%%%

\end{minipage}

\end{document}